\documentclass[aps, sort&compress, longtable, square, floatfix]{revtex4}
\usepackage[dvips]{graphicx}
\usepackage{times}

\begin{document}

\title{The Pedagogy of the p-n Junction: Diffusion or Drift?}

\author{S.O. Kasap}
\affiliation{Department of Electrical Engineering, University of Saskatchewan
Saskatoon S7N 0W0, Canada}

\author{C. Tannous}
\affiliation{Laboratoire de Magn\'etisme de Bretagne,  CNRS/UMR 6135,
6 avenue Le Gorgeu BP:  809,  29285 Brest CEDEX,  France}

\begin{abstract}
The majority of current textbooks on device physics at the undergraduate 
level derive the diode equation based on the diffusion of injected minority 
carriers. Generally the drift of the majority carriers, or the extent of 
drift, is not discussed and the importance of drift in the presence of a 
field in the neutral regions is almost totally ignored. The assumptions of 
zero field in the neutral regions and conduction by minority carrier 
diffusion lead to a number of pedagogical problems and paradoxes for the 
student. The purpose of this paper is to address the 
pedagogical problems and paradoxes apparent in the current treatment of 
conduction in the pn junction as it appears in the majority of texts.
\end{abstract}

\maketitle

\section{INTRODUCTION AND BACKGROUND}

The pn junction theory forms an integral part all physical electronics 
courses. At both undergraduate and graduate levels, the conventional 
analysis of the pn junction device under forward bias conditions follows 
closely Shockley's original treatment [1] in which the diode equation is 
derived based on the injection and diffusion of minority carriers. There 
are, however, a number of paradoxes in the treatment of the subject matter 
in the present textbooks (see below) which tends to mislead the students. 
Under zero and forward bias conditions, the pn junction displays a number
of characteristic features with particular reference 
to carrier concentrations, exposed ionized dopants in the space charge layer 
(SCL) or the depletion layer, internal field, $E(x)$ in the SCL, and the 
built-in voltage V$_{0}$. In a typical undergraduate course the arguments 
used in driving the diode equation follow a sequence of simplifying assumptions:

(a) The depletion region or the SCL (space charge layer) has a much higher resistance than the 
neutral regions  so that the applied voltage 
drops across the depletion region. There is therefore no field in the 
neutral regions. 

(b) The applied forward bias reduces the built-in potential V$_{0}$ and allows the 
diffusion and hence injection of minority carriers. 
From assumption (a), the law of the junction gives the injected minority 
carrier concentration in terms of the applied voltage. For example for holes 
injected into the n-side,

\begin{equation}
{\rm{p}}_{\rm{n}} {\rm{(0)}} = {\rm{p}}_{{\rm{n0}}} {\rm{exp(}}\frac{{{\rm{eV}}}}{{{\rm{kT}}}}{\rm{)}}
\label{eq1}
\end{equation}

where p$_{n}$(0) is the hole concentration just outside the depletion 
region, at the origin x=0, p$_{n0}$ is the equilibrium hole 
concentration in the n-region, p$_{n0}$=n$_{i}^{2}$/N$_{d}$, V is the 
applied bias, and the other symbols have their usual meanings.

(c) Since the electric field in the neutral regions is assumed to be zero, 
the continuity equation for the minority carriers is greatly simplified and 
becomes analytically tractable even at the junior undergraduate level. For 
holes in the n-region, under steady state conditions, $\partial 
$p$_{n}$/$\partial $t=0 and the continuity equation is simply

\begin{equation}
 -  \frac{{\rm{1}}}{{\rm{e}}}\frac{{{\rm{dJ}}_{{\rm{hn}}} }}{{{\rm{dx}}}}  - 
 \frac{{{\rm{p}}_{\rm{n}} }}{{\tau _{{\rm{hn}}} }} = 0
\label{eq2}
\end{equation}

where J$_{hn}$ is the hole current density and $\tau _{hn}$ is the 
minority carrier (hole) recombination time both in the n-region. It is 
tacitly assumed in eq.(\ref{eq2}) that the minority carrier concentration is much 
less than the equilibrium majority carrier concentration so that a constant 
minority carrier lifetime can be defined which is independent of the 
majority carrier concentration, n$_{n}$. 

The hole current density however is simply the diffusion component as the 
electric field is assumed to be zero,

\begin{equation}
J_{hn}  =  - eD_{hn} \frac{{d{\rm{p}}_{\rm{n}} }}{{{\rm{dx}}}}
\label{eq3}
\end{equation}

where D$_{hn}$ is the diffusion coefficient of the minority carriers (holes) 
in the n-region. 

Substituting eq. (\ref{eq3}) into eq. (\ref{eq2}), leads to

\begin{equation}
D_{hn} \frac{{{\rm{d}}^{\rm{2}} {\rm{p}}_{\rm{n}} }}{{{\rm{dx}}^{\rm{2}} }} 
 -  \frac{{{\rm{p}}_{\rm{n}} }}{{\tau _{{\rm{hn}}} }} = 0
\label{eq4}
\end{equation}

and solving eq.(\ref{eq4}) for p$_{n}$ one obtains, for a long diode,

\begin{equation}
\Delta {\rm{p}}_{\rm{n}} {\rm{(x)}} = \Delta {\rm{p}}_{\rm{n}} 
{\rm{(0)exp(-}}\frac{{\rm{x}}}{{{\rm{L}}_{{\rm{hn}}} }}{\rm{)}}
\label{eq5}
\end{equation}

where $\Delta $p$_{n}$=p$_{n}$-p$_{n0}$ is the excess minority carrier 
concentration and x is measured from just outside the depletion region. 
The long diode assumption means that the length of the 
neutral region, l$_{n}$, is much greater than the minority carrier diffusion 
length, L$_{hn}$, defined as $\sqrt{D_{hn} \tau_{hn}}$. The long 
diode assumption however is not necessary for the derivation of the diode 
equation; it serves to simply the solution of eq.(\ref{eq4}) to a single 
exponential rather than a hyperbolic sine function.

With the minority carrier concentration given as in eq.(\ref{eq5}), from eq.(\ref{eq3}) 
the hole current density is

\begin{equation}
J_{hn} {\rm{(x)}} = \frac{{eD_{hn} \Delta {\rm{p}}_{\rm{n}} (0)}}{{{\rm{L}}_{{\rm{hn}}} }}{\rm{exp(-}}\frac{{\rm{x}}}{{{\rm{L}}_{{\rm{hn}}} }}{\rm{)}}
\label{eq6}
\end{equation}

There is obviously a similar minority carrier diffusion current density in 
the p-region for the injected electrons, i.e.

\begin{equation}
J_{ep} {\rm{(x)}} = \frac{{eD_{ep} \Delta {\rm{n}}_{\rm{p}} (0)}}{{{\rm{L}}_{{\rm{ep}}} }}{\rm{exp(-}}\frac{{{\rm{x}}}}{{{\rm{L}}_{{\rm{ep}}} }}{\rm{)}}
\label{eq7}
\end{equation}

where D$_{ep}$ and L$_{ep}$ are the electron diffusion coefficient and 
diffusion length, $\Delta $n$_{p}$ is the excess electrons concentration all 
in the p-region, and x' is distance measured away from the depletion region 
in the p-side. 

(d) It is assumed that the depletion region is so narrow that the currents 
do not vary across this region. Then the majority carrier current at x=0 is 
the same as the minority carrier current at x'=0. Similarly the majority 
carrier current at x'=0 is the same as the minority carrier current at x=0. 
Thus the total current density is

\begin{equation}
J = J_{hn} (0) + J_{ep} (0)
\label{eq8}
\end{equation}

and using the law of the junction for $\Delta $p$_{n}$(0) and $\Delta 
$n$_{p}$(0) from eq.\ref{eq1} one obtains the diode equation,

\begin{equation}
J = \frac{{eD_{hn} \Delta {\rm{p}}_{\rm{n}} (0)}}{{{\rm{L}}_{{\rm{hn}}} }} 
+ \frac{{eD_{ep} \Delta {\rm{n}}_{\rm{p}} (0)}}{{{\rm{L}}_{{\rm{ep}}} }}
\label{eq9}
\end{equation}

or

\begin{equation}
J = e\left[ {\frac{{D_{hn} p_{n0} }}{{{\rm{L}}_{{\rm{hn}}} }} + 
\frac{{D_{ep} n_{p0} }}{{{\rm{L}}_{{\rm{ep}}} }}} \right] 
\left( {{\rm{exp}}(\frac{{eV}}{{kT}})  -  1} \right)
\label{eq10}
\end{equation}

This is the general diode equation found in the majority of textbooks which 
follow the above sequence of steps, either through tacitly or explicitly 
stated assumptions in (a) to (d). Many texts, for simplicity, consider 
either the p$^{ + }$n or the n$^{ + }$p junction. For the p$^{ + }$n 
junction, $N_{A} >> N_{D}$ and  eq.(11) becomes,

\begin{equation}
J = en_i^2 \left[ {\frac{{D_{hn} }}{{{\rm{L}}_{{\rm{hn}}} N_D }} + 
\frac{{D_{ep} }}{{{\rm{L}}_{{\rm{ep}}} N_A }}} \right] 
\left( {{\rm{exp}}(\frac{{eV}}{{kT}})  -  1} \right)
\label{eq11}
\end{equation}

The assumption stated explicitly in (d) is commonly overlooked in many 
current textbooks on the subject and is one of the key assumptions in the 
derivation of equation (\ref{eq2}) as discussed, for example, by Moloney \cite{moloney}.

Equation (\ref{eq2}) gives the impression that the conduction process in the
 p$^{ + }$n junction diode is the diffusion of injected holes in the n-region. 
The above steps invariably lead to a conclusion for the student that it is the 
minority carrier diffusion which constitutes the forward current.

Close examination of the above steps in the derivation exposes a number of 
serious pedagogical paradoxes and problems for the student and the 
instructor. The diode equation in eq. \ref{eq11} is so entrenched in our teaching 
of the pn junction that it has been used to design many simple but fruitful 
laboratory experiments as reported in various journals.

\section{PEDAGOGICAL PROBLEMS}

The conventional undergraduate level treatment in Section I leads to a 
number of pedagogical problems and paradoxes. We cite those we have 
encountered frequently in two undergraduate classes during the treatment of 
the forward biased long diode:

(a) If there is no field in the neutral region then there can be no net 
charge at any point in this region inasmuch as $dE/dx=0$. Then the excess 
majority carrier concentration should follow the decay of the excess 
minority carrier concentration, $\Delta $n$_{n}$(x)=$\Delta $p$_{n}$(x), 
which then follows eq. \ref{eq5}. But the gradients of $\Delta 
$n$_{n}$(x)=$\Delta $p$_{n}$(x) along x must be the same as well. Therefore 
majority carriers diffuse towards the right as well and since in silicon 
$D_{e} > D_{h} $, and electrons are negatively charged the net current is 
actually in the reverse direction. The current must be in the opposite 
direction to the applied voltage!

(b) The minority carrier current, which is due to diffusion, decays with x. 
Since the total current must be constant, the majority carrier current must 
increase with x. However, there is no field in the neutral region which 
means that electrons must diffuse. From the first paradox above this 
diffusion can not make up for the decay in the hole current. 

(c) Far away from the depletion region, both the hole and electron 
concentrations are almost uniform. If there is no electric field, then the 
current, due to diffusion, must vanish. How is it that the current stays 
constant in the neutral region (indeed in the whole device)?

(d) The absence of an electric field in the neutral region means that 
$\Delta $n$_{n}$(x)=$\Delta $p$_{n}$(x). But when holes are injected into 
the n-region they recombine with electrons so that , intuitively, the 
majority carrier concentration should decrease not increase.

The concept of field free neutral regions is so deeply rooted in the present 
treatment that many authors do indeed show the excess majority carrier 
concentration increasing towards the junction like the excess minority 
carrier concentration. This fact alone is contrary to student intuition 
that, if anything, the excess majority concentration should remain uniform. 
It is clear that the instructor has a responsibility to clear this paradox. 
It is interesting to note that a large number of authors sketch the carrier 
concentration profiles with the majority carrier concentration shown as 
uniform whereas others show the excess majority carrier profile following 
the excess minority carrier concentration profile as indicated in Table I 
for a survey of a large number of books on the subject. The differences in 
the diagrams only add confusion to the student's understanding. Few authors 
allude to the presence of the electric field and majority carrier drift to 
overcome some of the problems listed above. There is however no satisfactory 
treatment in the majority of the textbooks used in English speaking 
countries we have examined as illustrated in Table I. In most books the 
field in the neutral region is totally neglected in the treatment. The 
result, we believe, is pedagogical paradoxes and a student who is confused. 
It seems that only some of the early books published in the sixties 
consider the need and the effects of the field in the neutral regions in 
their discussion of conduction in the forward biased pn junction. 

\section{DIFFUSION OR DRIFT?}

For simplicity we consider a long p$^{ + }$n junction diode. Equation (\ref{eq2}) 
describes its conventional current-voltage characteristics. We use the 
parameters listed in Table 2 which represent "typical" parameters for a p$^{ 
+ }$n junction Si diode albeit classroom parameters. We also assume small 
injection so that $p_{n0} << n_{n0}$ (or N$_{D})$. The latter assumption 
means that the minority carrier recombination time remains a useful 
parameter . For forward bias we take the voltage across the diode to be 
typically 0.55V. The depletion region extends essentially into the n-side 
and its width W, is much shorter than the hole diffusion length in this 
region as indicated in Table II. Similarly the relatively tiny extension of 
the depletion width into the p$^{ + }$region is much shorter than the 
electron diffusion length there. Lengths of the neutral regions are taken to 
be about ten times their minority carrier diffusion lengths to represent a 
"long-diode", i.e. l$_{n}$=10L$_{h}$ and l$_{p}$=10L$_{e}$.

\begin{equation}
J = \frac{{eD_{hn} n_i^2 }}{{{\rm{L}}_{{\rm{hn}}} N_D }}\left( 
{{\rm{exp}}(\frac{{eV}}{{kT}})  -  1} \right)
\label{eq12}
\end{equation}

The first attempt to overcome the problems listed in Section II is to allow 
some of the applied voltage, a small fraction of it, to drop across the 
neutral n-region of the p$^{ + }$-n junction. This is easily accepted by the 
student since the neutral regions must have some finite resistance even 
though much smaller than the depletion region. This means that the law of 
the junction remains approximately valid. What is the field in the n-region?

The total current through the p$^{ + }$n junction must be continuous. This 
means that at any point in the n-region,

\begin{equation}
J =  - eD_{hn} \frac{{d\Delta p_n }}{{dx}} + e\mu _{hn} p_n E_n  + 
eD_{en} \frac{{d\Delta n_n }}{{dx}} + e\mu _{en} n_n E_n =  constant   \\
\label{eq13}
\end{equation}

where E$_{n}$ is the field in the n-region at x. Initially E$_{n}$ is 
assumed to be small but finite. 

Since $n_{n} >> p_{n}$ , $n_{n} \approx N_{D}$ (small injection), and 
$\Delta $n$_{n}$(x)=$\Delta $p$_{n}$(x) the above equation simplifies to,

\begin{equation}
J = e(D_{en}  - D_{hn} )\frac{{d\Delta p_n }}{{dx}} + e\mu _{en} N_D E_n(x) 
\label{eq14}
\end{equation}

The requirement of an internal field is quite transparent from eq. (\ref{eq3}). 
The minority concentration gradient is negative and $D_{en} > D_{hn}$ which 
means that the first term in eq.(\ref{eq3}) is negative so that current is in the 
negative direction. Unless there is an internal field drifting the majority 
carriers it is not possible to obtain a positive current.

The pedagogic development at this point must make use of the excess minority 
carrier concentration in eq.\ref{eq5}. If the field is indeed sufficiently small 
it may be assumed that the excess minority carrier profile, $\Delta 
$p$_{n}$(x), is still given by eq.\ref{eq5}. The validity of this assumption will 
be demonstrated below with an illustrative example. One can also assume that 
eq. (\ref{eq2}) can still be used to describe the total diode current. Then from 
eqs. \ref{eq5}, \ref{eq2} and \ref{eq3}, one can obtain the field E$_{n}$ in the n-region

\begin{equation}
E_n(x) = \left( {\frac{n_i }{N_D }} \right)^2\left( {\frac{kT}{eL_{hn} }} 
\right)\frac{1}{b_n }\exp (\frac{eV}{kT})\left[ {1 + (b_n - 1)\exp ( - 
\frac{x}{L_{hn} })} \right]
\label{eq15}
\end{equation}

where we have used the definition b$_{n}=\mu _{en}$/$\mu _{hn}$=D$_{en}$/D$_{hn}(>1) $.
 Equation (\ref{eq4}) describes the field outside 
the SCL in the so-called neutral region that is needed to maintain the diode 
current. In the n-region the field increases towards the SCL. The increase in the field is 
required to make up for the negative electron diffusion current. Far away 
from the depletion region the current is maintained by a constant field of 
magnitude,

\begin{equation}
E_{n}(x)  = \left( {\frac{{n_i }}{{N_D }}} \right)^2 \left( {\frac{{kT}}{{eL_{hn} }}}
 \right)\frac{1}{{b_n }}\exp (\frac{{eV}}{{kT}})\left[ {1 +
 (b_n  - 1)\exp ( - \frac{x}{{L_{hn} }})} \right]
\label{eq16}
\end{equation}

An interesting feature is that the magnitude of the field increases 
exponentially with the applied voltage contrary to student intuition based 
on the applied voltage simply dividing between the resistance of the 
depletion region and the resistance of the neutral region.

With the field given in eq.\ref{eq15} a paradox mentioned in Section II develops 
in that E varies spatially across the neutral region so that $dE/dx$ is not 
zero. Gauss equation in point form (or the Poisson equation) in the 
n-region is:

\begin{equation}
E_{n}(x \mapsto \infty)  = \left( {\frac{{n_i }}{{N_D }}} \right)^2 
\left( {\frac{{kT}}{{eL_{hn} }}} \right)\frac{1}{{b_n }}\exp (\frac{{eV}}{{kT}})
\label{eq17}
\end{equation}

\begin{equation}
\frac{\varepsilon }{e}\frac{{dE_n }}{{dx}} = \Delta p_n (x)-\Delta n_n (x)
\label{eq18}
\end{equation}

where $\varepsilon $ is the total permittivity of Si (=$\varepsilon 
_{o}\varepsilon _{r})$.

Since $\Delta $p$_{n}$(x) is determined by eq.\ref{eq5}, the excess majority 
carrier concentration is:

\begin{equation}
\Delta n_n (x) = \Delta p_n (x)\left[ {1 - \frac{{\frac{\varepsilon }{e}
\frac{{dE_n }}{{dx}}}}{{\Delta p_n (x)}}} \right] = \Delta p_n (x)
\left[ {1 + \frac{{\varepsilon (b_n  - 1)kT}}{{e^2 b_n N_D L_{hn}^2 }}} \right]
\label{eq19}
\end{equation}

Substituting typical values for $\varepsilon $, b$_{n}$, N$_{D}$, L$_{hn}$ 
from Table II into eq.(\ref{eq2}) shows that the second term is $\sim $4.1x10$^{ - 
6}$. Thus $\Delta $n$_{n}$(x)=$\Delta $p$_{n}$(x) and the charge neutrality 
condition for all practical purposes remains valid. We have found the 
requirement of $\Delta $n$_{n}$(x)=$\Delta $p$_{n}$(x) to be somewhat 
contrary to student intuition. This is further exasperated by many texts 
showing the majority carrier concentration uniform in the neutral regions 
(see Table I) which misleads the student. Qualitatively, the injected holes 
into the n-region disturb the charge neutrality and set-up a field here 
which then drives the electrons towards the SCL until a steady state is 
reached between electron drift and diffusion. Thus the increase in the 
majority carrier concentration towards the SCL is due to the driving effect 
of the field, E$_{n}$, even though it appears at first that $\Delta $n$_{n}$ 
should decrease towards theSCL as injected holes recombine with electrons 

Once the field in the n-region is given as eq.\ref{eq15}, the student can readily 
calculate the various contributions to the total current density using 
eq.\ref{eq13}. The magnitudes of the various current components (majority carrier diffusion
 and drift, and minority carrier diffusion and 
drift) and their directions 
are listed in Table II. In general, the \textit{drift of 
the majority carriers is the most significant contribution to the pn junction
diode current.} How is then the 
diffusion terminology comes to appear in explaining the diode current even 
though the biggest contribution is drift? 

Given that the depletion region is very narrow and that recombination in 
this region is negligibly small due to the very small concentrations of 
carriers, then in the steady state one must have 
$\frac{\partial J_e }{\partial x} = 0$ and the electron current J$_{e}$ 
must be constant through 
the SCL. But, electron drift at x'=0, i.e minority carrier drift, is 
negligible and the electron current there is primarily a diffusion current 
just like the hole current in the n-region. Thus the total electron current 
at x=0 must equal to the electron diffusion current at x'=0. Similarly the 
hole diffusion current at x=0 is equal to the total hole current at x'=0 
which is essentially by drift. It is apparent that by evaluating the 
\textit{minority carrier diffusion current just outside depletion region we 
are indirectly determining the total majority carrier current at the other 
side of the depletion region}. This is a subtle point that seems to have 
been short circuited in the 
majority of the texts. Consequently, the diode equation stated in equation 
(10) is only valid if the SCL width is much shorter than the minority 
carrier diffusion length in that region

An important paradox that must be addressed by the instructor is the much 
cherished minority carrier concentration profile stated in eq.\ref{eq5} for a 
long diode. Equation \ref{eq5} is the solution of the continuity equation in the 
absence of an electric field. As a first step one can assume a constant 
field, E$_{n}$, in the neutral region to examine its effects on the excess 
minority carrier concentration profile. In the presence of a constant field, 
the general continuity equation in eq.\ref{eq2} leads to 

\begin{equation}
D_{hn} \frac{{{\rm{d}}^{\rm{2}} {\rm{p}}_{\rm{n}} }}{{{\rm{dx}}^{\rm{2}} }} - 
 \mu _{hn} E_n \frac{{{\rm{dp}}_{\rm{n}} }}{{{\rm{dx}}}} - 
 \frac{{{\rm{p}}_{\rm{n}} }}{{\tau _{{\rm{hn}}} }} = 0
\label{eq20}
\end{equation}

Since this is a linear differential equation, the undergraduate student can 
readily solve it or accept its solution by substituting the solution into 
eq.(\ref{eq3}). For a long neutral region, the solution is:

\begin{equation}
\Delta {\rm{p}}_{\rm{n}} {\rm{(x)}} = \Delta {\rm{p}}_{\rm{n}} 
{\rm{(0)exp(-}}\frac{{\rm{x}}}{{{\rm{a}}_{\rm{n}} }}{\rm{)}}
\label{eq21}
\end{equation}

where a$_{n}$ is a "length constant" which can be determined by substituting 
eq.(\ref{eq4}) into (\ref{eq3});

\begin{equation}
{\rm{a}}_{\rm{n}}^{\rm{2}}  -  (\tau _{{\rm{hn}}} \mu _{hn} 
E_n {\rm{)a}}_{\rm{n}}  -  (D_{hn} \tau _{{\rm{hn}}} ) = 0
\label{eq22}
\end{equation}

Solving this quadratic equation we find,

\begin{equation}
{\rm{a}}_{\rm{n}}  =  \sqrt {(D_{hn} \tau _{{\rm{hn}}} )} 
\left[ {\sqrt {\alpha _n  + 1}  + \alpha _n } \right] = 
L_{hn} \left[ {\sqrt {\alpha _n  + 1}  + \alpha _n } \right]
\label{eq23}
\end{equation}

where $\alpha _{n}$ is defined as

\begin{equation}
\alpha _n  =  \frac{{\frac{1}{2}\tau _{{\rm{hn}}} 
\mu _{hn} E_n }}{{L_{hn} }} = 
\frac{{\frac{1}{2}L_{drift} }}{{L_{hn} }} = 
\frac{1}{2}\frac{{Schubweg}}{{Diffusion \hspace{1mm} Length}}
\label{eq24}
\end{equation}

The parameter $\alpha_{n}$ represents the comparative effect of drift to 
diffusion since L$_{drift}=^{1}$/$_{2}\tau _{hn}\mu _{hn}$E$_{n}$ 
is the so-called Schubweg of the minority carriers, distance drifted before 
recombination. If the field is small $\alpha  \quad _{n }$will be small and in 
the limit of zero field, E$_{n} \sim 0$, $a_{n}\sim L_{hn}$ and the
 theory approaches the conventional zero-field treatment. 
At a forward bias of 0.55V, the field is maximum at x=0, and using this 
maximum value one finds $\alpha _{n}$=0.00106 and 
a$_{n}$=(1.0011)L$_{hn}$. At V=0.65V, a$_{n}$=(1.05)L$_{hn}$ and a$_{n}$ is 
still very close to L$_{hn }$(within 5{\%}) even though the injected hole 
concentration is now only ten times smaller than the equilibrium majority 
concentration which sets the limit of small injection. Although the solution 
in eq.(\ref{eq3}) does not apply when the field is non-uniform as in eq.\ref{eq15}, it 
does nonetheless provide convenient means for the student to examine the 
possible effect of the field on the excess minority carrier profile. 

The field in the p-region can be similarly derived. The total current in the 
p$^{ + }$ region is

\begin{equation}
J = eD_{ep} \frac{{d\Delta n_n }}{{dx}} + e\mu _{ep} n_p E_p  - 
eD_{hp} \frac{{d\Delta p_p }}{{dx}} + e\mu _{hp} p_p E_p  =  constant  \\
\label{eq25}
\end{equation}

Since the total current must be constant and assigned to be described by 
eq. \ref{eq12}, using the corresponding version of eq. \ref{eq5} for minority and 
majority carrier excess concentration in the p-region one can derive

\begin{equation}
E_p(x)  = \exp (\frac{{eV}}{{kT}})\left( {\frac{{n_i^2 }}{{N_D N_A }}} \right)\left( {\frac{{kT}}{{eL_{hn} }}} \right)\left( {\frac{{\mu _{hn} }}{{\mu _{hp} }}} \right)\left[ {1 - C_p \exp ( - \frac{{x'}}{{L_{ep} }})} \right]
\label{eq26}
\end{equation}

where C$_{p}$ is defined as

\begin{equation}
C_p  = \left( {b_p  - 1)} \right)\left( {\frac{{\mu _{hp} }}{{\mu _{hn} }}} \right)
\left( {\frac{{L_{hn} }}{{L_{ep} }}} \right)\left( {\frac{{N_D }}{{N_A }}} \right)
\label{eq27}
\end{equation}

in which b$_{p }=\mu _{ep}$/$\mu _{hp}$ is the electron to hole drift 
mobility ratio in the p-region. We assumed that, as usual under forward 
bias, $V>>kT/e $. Equation (\ref{eq9}) shows that the field is minimum right at the 
SCL, x'=0, and increases exponentially to a constant value away from the 
junction. Substituting typical values from Table 
II shows that $E_{p}<<E_{n}$. When the two fields are compared, one finds  
that E$_{n}$ is at least three orders of magntiude greater 
than E$_{p}$. In fact E$_{p}$ is almost unifrom in the p$^{ + }$-region. 
Most interestingly and importantly, even though E$_{p}$ is even smaller than 
E$_{n}$, its effect is most significant. One readily can calculate the 
contributions of each term to the current density in the p-region from eq. \ref{eq24}.
The values at the SCL are listed in Table II where it is apparent that 
the current is carried almost totally by \textit{majority carrier drift}. 
A distinctly different behavior 
in the p$^{ + }$-region from that observed in the n-region is the fact that 
the majority carrier diffusion is insignificant and that minority carrier 
diffusion, though larger than majority diffusion, is some three orders of 
magnitude smaller than majority carrier drift. From the above discussion for 
conduction on the n-side it is clear that in deriving the diode equation we 
are calculating the hole (majority) drift current in the p$^{ + }$-region by 
evaluating the hole (minority) diffusion current in the n-region simply 
because the total hole current does not change through the SCL as long as 
the latter is thinner than the hole diffusion length. 

It is always useful for the student to reconfirm that the majority of the 
voltage drops across the depletion region by evaluating the voltage drop 
across the neutral regions. If V$_{n}$ is the voltage drop across the 
n-region then

\begin{equation}
V_n  = \int_0^{l_n } {E_n dx} 
\label{eq28}
\end{equation}

or

\begin{equation}
V_n  = \left( {\frac{{n_i }}{{N_D }}} \right)^2 \left( {\frac{{kT}}{{eL_{hn} }}}
 \right)\frac{1}{{b_n }}\left( {l_n  + (b_n  - 1)L_{hn} } \right)\exp (\frac{{eV}}{{kT}})
\label{eq29}
\end{equation}

Equation (\ref{eq2}) shows that the voltage drops increases exponentially with the 
applied bias contrary to an intuitive guess. Using typical values, at 
V=0.55V, V$_{n}$ is 0.00168V, whereas at V=0.6V, V$_{n}$ is 0.019V and the 
injected hole concentration in the n-region is about 11{\%} of N$_{D}$ which 
is the limit of small injection. At V=0.65V, V$_{n}$ becomes 0.121V which is 
quite significant but at this bias voltage the injected hole concentration 
is no longer small compared with N$_{D}$. There is a clear indication that 
as the voltage across the diode increases more and more of the applied 
voltage drops across the neutral regions which deteriorates the law of the 
junction. It is not difficult to show that since $E_{p}<<E_{n}$ and 
$l_{p} << l_{n}$, the voltage drop across the p$^{ + }$-region is orders of 
magnitude smaller than V$_{n}$.

\section{NUMERICAL DESCRIPTION OF THE PN JUNCTION:}
While the previous sections discussed the Diffusion/Drift approximation problem,
this section is oriented towards an \textit{exact} description of the PN
junction without making any assumptions. \\  

Starting from the constitutive system of equations \cite{polak}:

\begin{equation}
\frac{d\psi }{dx} = - E
\end{equation}

\begin{equation}
J_n = e\left[ {D_n } \right.\frac{dn}{dx} - \mu _n n\frac{d\psi }{dx}\left. 
\right]
\end{equation}

\begin{equation}
J_p = - e\left[ {D_p } \right.\frac{dp}{dx} + \mu _p p\frac{d\psi 
}{dx}\left. \right]
\end{equation}

\begin{equation}
\frac{1}{e}\frac{dJ_n }{dx} = R(n,p) + G(x)
\end{equation}

\begin{equation}
\frac{1}{e}\frac{dJ_p }{dx} = - R(n,p) - G(x)
\end{equation}

with the recombination term of the Shockley-Read-Hall form:

\begin{equation}
R(n,p) = \frac{(np - n_i^2 )}{T_1 n + T_2 p + T_3 }
\end{equation}

where T$_{1}$, T$_{2 }$and T$_{3 }$are time constants. Typically for Silicon 
T$_{1}=\tau _{p}$= 10$^{ - 5}$ sec,$_{ }$T$_{2}=\tau _{n}$= 10$^{ 
- 5}$ sec$_{ }$and T$_{3}$= ($\tau _{p }+\tau _{n})$n$_{i}$ \cite{ringhofer}.

In order to solve the boundary value problem associated with the above 
system (when a voltage is applied to the PN junction) we transform it into a 
new hybrid system of first-order (Current and carrier density equations) and 
one second-order differential equation (Poisson equation).

The mathematical/numerical reasons for performing this transformation reside 
in the fact the above system is a "singularly singular perturbed problem" 
\cite{mark84,ascher}. Many algorithms \cite{ascher,ascher83} have been 
developed in order to deal with this difficulty stemming from several facts:
\begin{enumerate}

\item $\psi $, n and p are fast variables in comparison with E, J$_{n}$ and 
J$_{p}$ \cite{ringhofer,mock,please}.

\item Near the limits of the depletion layer the values of n and p change by 
several orders of magnitude making the space-charge zone a double boundary 
layer. This difficulty is of the same type as the one encountered in 
Hydro/Aerodynamics where the fluid velocity field changes by several orders 
of magnitude near an obstacle.

Recognizing the difficulty due to the presence of the space-charge layer, a 
standard way to find a valid solution is to treat the boundary layer 
separately from the rest of the diode. In spite of the success of this 
approach \cite{mari,arand}, one might feel uneasy about this artificial 
dichotomy and rather tackle the problem with new powerful mathematical/numerical
methods that handle the layer and the rest of the device on the same equal footing. 

\item When the above system is rewritten explicitly in terms of the Poisson 
equation as we will do below, the second spatial derivative of the electric 
potential is multiplied by a very small number $\lambda ^{2}$ 
($\lambda \sim 10^{-4} \mbox{to} 10^{-3})$. Actually, this is the reason the 
problem is called singularly perturbed: the solution with $\lambda=0$ 
is entirely different from the solution with $\lambda $ finite 
but small \cite{ascher}.

\end{enumerate}

One of the early algorithms aimed at circumventing the above difficulties is 
the Scharfetter-Gummel algorithm \cite{scharfetter}. The latter attempts at segregating 
the fast/slow variables by integrating out the fast variables over some 
small interval while holding the slow variables constant over that same 
interval. The Scharfetter-Gummel algorithm leads to a spatial exponential 
discretization that will alleviate for the rapid variation of the fast 
variables.

Many variants of the Scharfetter-Gummel algorithm have been developed \cite{he} 
in order to cure some of its shortcomings which generally are numerical 
oscillations and crosswind effects. These lead to a loss of accuracy of the 
solution and sometimes preclude convergence towards the solution.

We decided not to use the Gummel algorithm or any of its variants but rather 
to tackle the problem from the singular perturbation point of view since 
this approach is more rigorous and leads to a better understanding and 
control of the instability problem encountered in the semiconductor system of 
equations. 

We first transform the system in the following dimensionless two-point 
boundary value problem with no generation processes considered:

\[
\frac{dn}{dx} = C_1 J_n + n\frac{d\psi }{dx}
\]

\[
\frac{dp}{dx} = - C_2 J_p - p\frac{d\psi }{dx}
\]

\[
\frac{dJ_n }{dx} = C_3 \frac{(np - 1)}{n + \tau _1 p + \tau _2 }
\]

\[
\frac{dJ_p }{dx} = - C_3 \frac{(np - 1)}{n + \tau _1 p + \tau _2 }
\]

\[
\frac{d^2\psi }{dx^2} = C_4 (n - p + N_D )
\]

The constants C$_{1}$,$_{ }$C$_{2}$,$_{ }$C$_{3}$ and C$_{4}$ are given by:

\[
C_1 = \frac{J_0 L_D }{en_i D_n }, C_2 = \frac{J_0 L_D }{en_i D_p }, C_3 = \frac{e L_D n_i }{J_0 T_1 },  
C_4 = \frac{ e L_D^2 n_i }{\epsilon_s U_T }
\]

where the Debye length L$_{D}$ is given by $L_D=\sqrt{\frac{k_B T \epsilon_S}{n_i e^2}}$ and the scaling current
$J_0=\frac{n_i \mu_n k_B T}{L_D}$. 

The thermal voltage U$_{T}$=k$_{B}$T/e (T is the temperature and k$_{B}$ is 
Boltzmann constant). The time constants are now 
$\tau_{1}$=T$_{2}$/T$_{1}$ and $\tau_{2}$=T$_{3}$/(T$_{2}$n$_{i})$.

The above system is now in the appropriate form to integrate with a powerful 
B-spline collocation based algorithm specifically tailored for two point 
singularly perturbed boundary value problems: COLSYS \cite{ascher,mark86,ascher83}.
The algorithm is based on a controllable meshing technique \cite{ascher83,mark83} of the 
boundary layer which will lead essentially to the damping of the existing
singularities. The layer-damping mesh, being exponential in nature, 
encompasses the Scharfetter-Gummel case and can be shown rigorously to have 
the form:

\[
h_{i}=h_{i-1} \exp(\alpha h_{i-1}/\lambda )
\]

where $h_{i}$ is the i-th mesh point, $\alpha $ is a constant related to the 
required accuracy and the nature of the collocation and $\lambda {\rm g}$s 
the singular value parameter.

Previously, Markowich et al. \cite{mark84} tackled this problem from the same angle 
but they solved the symmetric diode case with one boundary-layer at the 
origin.

In this work we tackle the non-symmetric case where a double-boundary layer 
is present around the origin starting from very accurate initial conditions.

Varying the applied bias by steps of 0.1V we calculate the carrier density 
profiles, the potential and the electric field.

The drift and diffusion current density profiles are also obtained. Several 
tests are used in order to check the validity of the solution obtained. The 
first test is an accuracy test whereby we require a given accuracy and check 
whether the criterion is met. The next test is based on the requirement of 
convergence: the collocation builds a non-linear set of equations that has 
to be solved iteratively. The additional tests are the independant checks of 
the constancy of the current densities locally on each side of the junction 
and globally over the entire junction. The tests are shown in the current 
density profile figures.

The final test we use is the approximate validity of Shockley's equation. 
Varying the voltage, we obtain the IV characteristics of the junction and we 
compare it to the Shockley's case. Since we have used the Shockley-Read-Hall 
recombination term all over the PN junction we do not expect the Shockley's 
case but rather the general form:

\[
I=I_{S} [\exp(eV/\eta k_{B}T)-1]
\]

The comparison of the obtained IV characteristic to the Shockley formula is 
displayed in Fig. 1. The calculated characteristic falls between the two 
Shockley curves $\eta=1.1$ and  $\eta=1.2$ in a finite current interval.
This means, the  general Shockley formula is not valid, within the singular 
perturbation approach, for arbitrary current values.

\begin{figure}[!h]
\begin{center}
\scalebox{0.3}{\includegraphics[angle=-90]{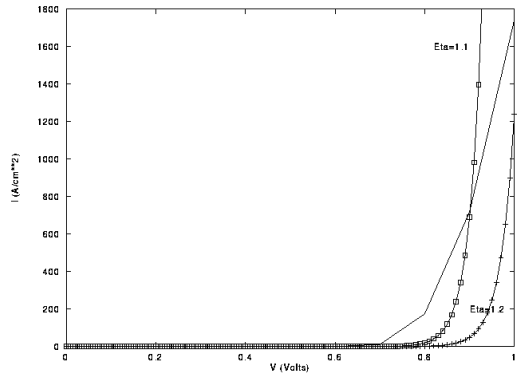}}
\end{center}
  \caption{IV characteristics obtained for ther PN junction and Shockley approximation
with  $\eta=1.1$ and  $\eta=1.2$.}
\label{fig1}
\end{figure}

\section{Conclusions}
The physics of the PN junction is gaining back interest with new developments in the
area of nanoelectronics specially in the area of spintronics where one has to account for 
the spin of the carriers in addition to their charge. The usual approximations that are valid
and successful in the description of the PN junction physics at the micron scale must be entirely
reviewed and adapted to the nano scale. The diffusion/drift approximation as well as the
nature of the singularities of the problem have been reviewed and reformulated in a way such that
the underlying assumptions are revealed with their consequences.

\newpage

\section{Tables}

\begin{table}[ht!]
\begin{tabular}
{|p{86pt}|p{50pt}|p{93pt}|p{86pt}|p{108pt}|p{108pt}|}
\hline
AUTHOR& 
REF.& 
SEMIQUANTITIVE DRIFT ANALYSIS& 
FIELD IN NEUTRAL REGIONS& 
MAJORITY CARRIER CONCENTRATION& 
COMMENT \\
\hline
Zambuto \cite{zambuto}& 
Ch. 6& 
Qualitative& 
Not discussed& 
Not shown& 
Undergraduate \\
\hline
Yang \cite{yang}& 
Ch. 4& 
Qualitative& 
Not discussed& 
$\Delta $n$_{n}$(x)=$\Delta $p$_{n}$(x)& 
Undergraduate \\
\hline
Elliott and Gibson \cite{elliott}& 
Ch. 9& 
None& 
Not discussed& 
n$_{n}$ shown constant& 
Undergraduate \\
\hline
Fraser (UK)\cite{fraser}& 
Ch. 3& 
None& 
Not discussed& 
n$_{n}$ shown constant& 
Undergraduate \\
\hline
Seymour \cite{seymour}& 
Ch. 3& 
None& 
Not discussed& 
Not shown& 
Undergraduate \\
\hline
Streetman \cite{streetman}& 
Ch. 5& 
Some (Example 5.4)& 
Mentioned but not discussed& 
Not shown& 
Undergraduate \\
\hline
Valdes \cite{valdes}& 
Ch. 9& 
Yes& 
Discussed& 
$\Delta $n$_{n}$(x)=$\Delta $p$_{n}$(x)& 
Out of print / Undergraduate \\
\hline
Sze \cite{sze85} & 
Ch. 3& 
None& 
Not discussed & 
Not clear & 
Undergraduate \\
\hline
Carroll (UK) \cite{carroll}& 
Ch. 4& 
None& 
Mentioned but not discussed & 
Not clear& 
Undergraduate \\
\hline
Solymar and Walsh (UK) \cite{solymar}& 
Ch.9& 
None& 
Not discussed& 
$\Delta $n$_{n}$(x)=$\Delta $p$_{n}$(x) implied& 
Undergraduate \\
\hline
Shur \cite{shur}& 
Ch. 2& 
Some (Fig. 2.3-8)& 
Not discussed& 
$\Delta $n$_{n}$(x)=$\Delta $p$_{n}$(x) implied& 
Senior UG / Graduate level \\
\hline
Pulfrey and Tarr (Canada) \cite{pulfrey}& 
Ch. 6& 
None& 
Not discussed& 
$\Delta $n$_{n}$(x)=$\Delta $p$_{n}$(x) implied but n$_{n}$=constant in diagrams& 
Undergraduate \\
\hline
Colclaser and Diehl-Nagle \cite{colclaser}& 
Ch. 7& 
None& 
Not discussed& 
Not mentioned and not shown& 
Undergraduate \\
\hline
Navon \cite{navon} & 
Ch. 6& 
None& 
Not discussed& 
$\Delta $n$_{n}$(x)=$\Delta $p$_{n}$(x) in diagram & 
Undergraduate \\
\hline
Sze \cite{sze81}& 
Ch. 2& 
None& 
Not discussed& 
Not clear& 
Senior Undergraduate / Graduate \\
\hline
Wang  \cite{swang}& 
Ch. 14& 
None& 
Not discussed& 
Inferred& 
Senior Undergraduate \\
\hline
Goodge (UK) \cite{goodge}& 
Ch.1& 
None& 
Not discussed& 
$\Delta $n$_{n}$(x)=$\Delta $p$_{n}$(x) in diagram& 
Undergraduate \\
\hline
Grove \cite{grove}  & 
Ch.3& 
None& 
Not discussed& 
Not clear & 
Out of print / Undergraduate \\
\hline
Van Der Ziel \cite{ziel}& 
Ch. 15& 
None& 
Not discussed& 
Not clear & 
Out of print / Undergraduate \\
\hline
Lonngren \cite{lonngren}& 
Ch. 6& 
None& 
Not discussed& 
Not discussed& 
Undergraduate \\
\hline
Tyagi \cite{tyagi}& 
Ch. 7& 
Qualitative& 
Mentioned but not discussed& 
$\Delta $n$_{n}$(x)=$\Delta $p$_{n}$(x) (Fig.7.3)& 
Undergraduate \\
\hline
Gibbons \cite{gibbons}& 
Ch. 6& 
Yes & 
Discussed& 
$\Delta $n$_{n}$(x)=$\Delta $p$_{n}$(x)& 
Out of print / Undergraduate \\
\hline
Ferendeci \cite{ferendeci}& 
Ch. 8& 
None& 
None& 
Not clear& 
Undergraduate \\
\hline
Allison \cite{allison}& 
Ch. 7& 
None& 
None& 
Not clear& 
Undergraduate \\
\hline
Neamen \cite{neamen}& 
Ch. 8& 
Some& 
E estimated& 
Not clear& 
Undergraduate \\
\hline
F. Wang \cite{fwang}& 
Ch. 8& 
None& 
None& 
Not clear& 
 \\
\hline
Gray et al. \cite{gray64}& 
Ch.2 and App. B. & 
Some& 
Yes& 
$\Delta $n$_{n}$(x)=$\Delta $p$_{n}$(x) in diagram& 
Out of print / Undergraduate \\
\hline
\end{tabular}
\caption{Treatment of the pn junction in a selection of the textbooks suitable for a 
physical electronics (solid state electronic devices) course.}
\end{table}
\label{tab1}

\squeezetable

\begin{table}
\begin{tabular}{|p{216pt}|p{194pt}|p{252pt}|}
\hline
\textbf{PROPERTY / PARAMETER}& 
\textbf{TYPICAL VALUE}& 
\textbf{COMMENT} \\
\hline
Permittivity& 
11.9& 
 \\
\hline
Intrinsic concentration, ni, cm$^{ - 3}$& 
1.5$\times $10$^{10}$& 
 \\
\hline
Donor concentration, N$_{D}$, cm$^{ - 3}$& 
5$\times $10$^{15}$& 
 \\
\hline
Acceptor concentration, N$_{A}$, cm$^{ - 3}$& 
10$^{19}$& 
p$^{ + }$-n junction \\
\hline
Equilibrium hole concentration \par in n-region, p$_{n0}$, cm$^{ - 3}$.& 
4.50$\times $10$^{4} $& 
 \\
\hline
Equilibrium electron concentration  \par in p-region, n$_{p0}$, cm$^{ - 3}$.& 
22.5 \quad & 
 \\
\hline
Hole recombination time in n-region, $\tau _{hn}$, s& 
$\tau _{hn} = \frac{5\times 10^{ - 7}}{1 + 2\times 10^{ - 17}N_D }$ \par $\tau _{hn} = 4.54\times 10^{ - 7}$& 
$\tau _{en }=\tau _{hn}$ \\
\hline
Electron recombination time in p-region, $\tau _{ep, }$s& 
$\tau _{ep} = \frac{5\times 10^{ - 7}}{1 + 2\times 10^{ - 17}N_A }$ \par $\tau _{ep} = 2.49\times 10^{ - 9}$& 
$\tau _{hp }=\tau _{ep}$ \\
\hline
Electron drift mobility in the n-region,  \par cm$^{2}$V$^{ - 1}$s$^{ - 1}$.& 
$\mu _{en} = 88 + \frac{1252}{1 + 6.984\times 10^{ - 18}N_D }$ \par $\mu _{en} = 1298$& 
$b_n = \frac{\mu _{en} }{\mu _{hn} } = 2.86$ \\
\hline
Hole drift mobility in n-region,  \par cm$^{2}$V$^{ - 1}$s$^{ - 1}$ &  $\mu _{hn}  = 54.3 + \frac{{407}}{{1 + 3.745 \times 10^{ - 18} N_D }} $ 
\par $\mu _{hn}  = 453.8$ & \\
\hline
Electron drift mobility in the p-region,  \par cm$^{2}$V$^{ - 1}$s$^{ - 1}$.& $\mu _{ep}  = 88 + \frac{{1252}}{{1 + 6.984 \times 10^{ - 18} N_A }}$
\par $\mu _{ep}  = 105.7$  & $b_p = \frac{\mu _{ep} }{\mu _{hp} } = 1.63$ \\
\hline
Hole drift mobility in p-region,  \par cm$^{2}$V$^{ - 1}$s$^{-1}$.&  $\mu _{hp}  = 54.3 + \frac{{407}}{{1 + 3.745 \times 10^{ - 18} N_A }}$
 \par $\mu _{hp}  = 64.9 $ &  \\
\hline
Electron diffusion coefficient in n-region, D$_{en}$, cm$^{2}$.s$^{ - 1}$&  33.55&   $ \frac{D_{en}}{\mu _{en}}=  \frac{kT}{e}$\\
\hline
Hole diffusion coefficient in n-region, D$_{hn}$,  \par cm$^{2}$.s$^{ - 1}$& 
11.73&  \\
\hline
Electron diffusion coefficient in p-region, D$_{ep}$, cm$^{2}$.s$^{ - 1}$& 
2.73& 
 \\
\hline
Hole diffusion coefficient in p-region, D$_{hp}$, cm$^{2}$.s$^{ - 1}$& 
1.68& 
 \\
\hline
Electron diffusion length in n-region, L$_{en}$, cm& 
3.90$\times $10$^{ - 3}$&   L$_{en}$=$\sqrt{D_{en}\tau_{en}}$
 \\
\hline
Hole diffusion length in n-region, L$_{hn}$, cm& 
2.31$\times $10$^{ - 3}$& 
 \\
\hline
Electron diffusion length in p-region, L$_{ep}$, cm& 
8.24$\times $10$^{ - 5}$ & 
 \\
\hline
Hole diffusion length in p-region, L$_{hp}$, cm& 
6.46$\times $10$^{ - 5}$& 
 \\
\hline
Builtin potential, V$_{bi}$, V& 
0.854& 
 \\
\hline
E$_{bi }$, V.cm$^{ - 1}$& 
3.60$\times $10$^{4}$& 
No bias \\
\hline
Width, W, of depletion region, cm& 
4.741$\times $10$^{ - 5}$& 
On n-side. Much shorter than hole diffusion length on n-side. \\
\hline
Width, W$_{n}$, of depletion region in n-side, cm& 
4.739$\times $10$^{ - 5}$& 
Much shorter than hole diffusion length in n-region. \\
\hline
Width, W$_{p}$ of depletion region in p-side, cm& 
2.369$\times $10$^{ - 8}$& 
Much shorter than electron diffusion length in p-region \\
\hline
Length of n-side& 
2.31$\times $10$^{ - 2}$& 
10L$_{h}$ Long diode \\
\hline
Length of p$^{ + }$-side& 
8.24$\times $10$^{ - 4}$& 
10L$_{e }$Long diode \\
\hline
\textbf{FORWARD BIAS, V}& 
\textbf{0.55}& \\
\hline
Built-in electric field, V.cm$^{ - 1}$& 
2.15$\times $10$^{4}$& 
Smaller than zero bias case \\
\hline
Width, W, of depletion region at 0.55V, cm& 
2.83$\times $10$^{ - 5}$& 
On n-side. Narrower under forward bias. Much shorter than hole diffusion length on n-side. \\
\hline
p$_{n}$(0), injected hole concentration at x=0& 
7.81$\times $10$^{13}$& 
 \\
\hline
n$_{p}$(0), injected electron concentration at x'=0& 
3.91$\times $10$^{10}$& 
 \\
\hline
p$_{n}$(0)/N$_{D}$& 
0.0156& 
1.56{\%}, small injection. At V=0.60V, this becomes 11{\%} \\
\hline
n$_{p}$(0)/N$_{A}$& 
3.91$\times $10$^{ - 9}$& 
Extremely small injection \\
\hline
J$_{0h}$, A.cm$^{ - 2}$& 
3.663$\times $10$^{ - 11}$& 
 \\
\hline
J$_{0e}$, A.cm$^{ - 2}$& 
1.195$\times $10$^{ - 13}$& 
  $J_{0e} << J_{0h}$ \\
\hline
J$_{0}$, A.cm$^{ - 2}$& 
3.675$\times $10$^{ - 13}$& 
$ \approx $J$_{0h}$ \\
\hline
J$_{0.55}$, A.cm$^{ - 2}$& 
0.0638& 
$J=J_{0}[\exp(V/kT)-1]$ \\
\hline
E$_{n\times }$, V.cm$^{ - 1}$& 
0.0613& 
Field in n-region far away from junction \\
\hline
E$_{nmax}$, V.cm$^{ - 1}$& 
0.1749& 
Field just outside SCL \\
\hline
E$_{nmax}$/E$_{bi }$& 
8.13$\times $10$^{ - 6}$& 
 \\
\hline
V$_{n }$at V=0.55V& 
0.00168& 
Very small, V$_{n }$is 0.3{\%} of bias \\
\hline
V$_{n }$at V=0.60V& 
0.0116& 
V$_{n }$is 12.3{\%} of bias  \\
\hline
E$_{pmax}$ = E$_{p\times }$, V.cm$^{ - 1}$& 
6.1$\times $10$^{ - 4}$& 
Field far away from junction \\
\hline
E$_{pmax}$/E$_{bi }$& 
2.8$\times $10$^{ - 8}$& 
Extremely small \\
\hline
V$_{p}$& 
5.04$\times $10$^{ - 7}$& 
Extremely small, $ V_{p}<< V_{n} << V $ \\
\hline
\textbf{JUST OUTSIDE SCL ON n-SIDE, x=0}& & \\
\hline
Majority drift current, A.cm$^{ - 2}$& 
0.1820 & 
Largest magnitude in positive direction \\
\hline
Majority diffusion current, A.cm$^{ - 2}$& 
-0.1818& 
Opposite direction, about the same magnitude as majority drift \\
\hline
Minority drift current, A.cm$^{ - 2}$& 
0.001& 
Smallest magnitude \\
\hline
Minority diffusion current, A.cm$^{ - 2}$& 
0.06359& 
About a third of the magnitude of majority drift. \\
\hline
\textbf{JUST OUTSIDE SCL ON p-SIDE, x'=0}& & \\
\hline
Majority drift current, A.cm$^{ - 2}$& 
0.06351& 
Largest magnitude. Dominates conduction. \\
\hline
Majority diffusion current, A.cm$^{ - 2}$& 
-0.000891& 
Opposite direction \\
\hline
Minority drift current, A.cm$^{ - 2}$& 
2.33$\times $10$^{-19}$& 
Smallest magnitude-virtually zero \\
\hline
Minority diffusion current, A.cm$^{ - 2}$& 
0.00255& 
Next largest magnitude \\
\hline
\end{tabular}
\caption{Properties of the p$^{ + }$-n junction }
\label{tab2}
\end{table}

\end{document}